\documentclass[fleqn,usenatbib]{mnras}

\usepackage{newtxtext,newtxmath}

\usepackage[T1]{fontenc}
\usepackage{ae,aecompl}

\usepackage{graphicx}	
\usepackage{amsmath}	
\usepackage{amssymb}	
\usepackage{verbatim}

\numberwithin{equation}{section}

\newcommand{\kms}{\mathrm{km~s}^{-1}}
\newcommand{\Msol}{\mathrm{M}_{\odot}}

\title[GalactICS with Gas]{GalactICS with Gas}

\author[Deg et al.]{{N. Deg$^{1}$\thanks{E-mail:nathan.deg@ast.uct.ac.za}, 
L. M. Widrow$^{2}$, T. Randriamampandry$^{1,3}$, and C. Carignan$^{1,4}$}\\
$^{1}$Department of Astronomy, University of Cape Town, Private Bag X3, Rondebosch 7701, South Africa \\                
$^{2}$Department of Physics, Engineering Physics, and Astronomy, Queen's University, Kingston, ON, K7L 3N6, Canada\\
$^{3}$ Kavli Institute for Astronomy and Astrophysics, Peking University, Beijing 100871, China \\
$^{4}$Observatoire d'Astrophysique de l'Universite de Ouagadougou (ODAUO), BP 7021, Ouagadougou 03, Burkina Faso}

\date{Accepted XXX. Received YYY; in original form ZZZ}

\pubyear{2018}

\begin{document}
\label{firstpage}
\pagerange{\pageref{firstpage}--\pageref{lastpage}}
\maketitle
\begin{abstract}
We present a new version of the \textsc{GalactICS} code that can generate self-consistent equilibrium 
galaxy models with a two-component stellar \textcolor{black}{disc} and a gas \textcolor{black}{disc} as well
as a centrally-concentrated bulge and extended dark halo. The models can serve as initial conditions for 
simulations of isolated galaxies that include both hydrodynamics and collisionless dynamics.
We test the code by evolving a pair of simple gas \textcolor{black}{disc}-halo models,
which differ only in the initial temperature
of the gas component. The models are similar to the ones considered in the Wang et al. except that here, the halo is live whereas they included the halo
as a fixed potential.
\textcolor{black}{We find that the basic structural properties of the models, such as the rotation curve and surface density profiles,
are well-preserved over 1.5 Gyr.} We also construct a Milky Way model that includes
thin and thick stellar \textcolor{black}{disc} components, a gas \textcolor{black}{disc}, a bulge, and a dark halo. Bar formation
occurs in all \textcolor{black}{disc}-like components at about \textcolor{black}{$1\,{\rm Gyr}$}. The bar 
is strongest in the thin \textcolor{black}{disc} while the gas \textcolor{black}{disc} contains the most
prominent spiral features. The length of the bar in our model is comparable to what has been inferred
for the Galactic bar.
\end{abstract}




\section{Introduction}

A fundamental problem in galactic astronomy is the 
construction of self-consistent equilibrium models 
for individual galaxies. Such models can be used as a 
template for interpreting observations and inferring the gravitational potential
of a galaxy and the structure of its dark halo \citep{Widrow2008, Taranu2017}.  
Furthermore, equilibrium models can 
provide initial conditions (ICs) for N-body simulations,
which can in turn be used to study dynamical processes
such as the formation of bars and spiral structure \citep{Fux1997,Athanassoula2002}.
Not only are these processes interesting in and of themselves, they
also provide additional constraints on the models.  
For example, an equilibrium model that satisfies observational
constraints but is nevertheless unstable to the formation of a strong
bar is not a suitable model for \textcolor{black}{a galaxy that has a weak bar or no bar at all}
\citep{Sellwood1985,Widrow2008,Toky2018}.

It is invariably easier to model the 
evolution of collisionless matter (stars and dark matter) than
gas.  The dynamics of collisionless matter is determined solely 
by the gravitational field, \textcolor{black}{which can be calculated efficiently using
particle mesh or 
tree codes}.  On the other 
hand, the dynamics of collisional matter depends on both the 
gravitational field and additional physics
including hydrodynamical forces, star formation, feedback, 
and turbulence. Furthermore, many of these processes 
occur at scales below the resolution limit of simulations.

The complications associated with gas components extend to setting up ICs,
especially, if one aims to follow a system from some equilibrium (but 
possibly unstable) state. ICs for collisionless systems in general require a self-consistent model for the phase space distribution 
function (DF) and the gravitational potential.
Collisional systems also require models
for the temperature of the gas as a function of position as well as the equation of state.

\citet{Springel2005b} described a method for initializing a system with gas and stellar \textcolor{black}{discs} 
as well as a dark halo. The DF for the collisionless components are found by solving the 
Jeans equations (see \citep{Hernquist1993} and discussion below). For the gas \textcolor{black}{disc}, they assumed a
simple equation of state $P=P(\rho)$. The vertical structure is then
determined by solving the equations for hydrostatic equilibrium \textcolor{black}{in the 
direction normal to the \textcolor{black}{disc} plane}.

This method was refined \citet{Wang2010}, who were particularly interested in
initializing \textcolor{black}{discs} in their adaptive mesh refinement (AMR) simulations.
In brief, they use an iterative approach to solve a set of equations describing 
the potential, surface density, and scale height of an 
isothermal, approximately exponential \textcolor{black}{disc} in the presence of 
an external potential. However, their simulations
were limited to a gas \textcolor{black}{disc} in a fixed potential.
In this paper we describe how to combine the
\textsc{GalactICS} (Galaxy Initial ConditionS) 
code \citep{Kuijken1995,Widrow2005,Widrow2008}, which 
generates collisionless bulge-\textcolor{black}{disc}-halo systems,
with the \citet{Wang2010} method of generating 
gas \textcolor{black}{discs}\footnote{The new version of \textsc{GalactICS} 
is available upon request.}.

The problem of generating equilibrium ICs for collisionless systems, though more straightforward
than that for collisional ones, is still a non-trivial 
task. The problem amounts to finding the DF for the
collisionless components that satisfy (at least approximately)
the coupled collisionless Boltzmann and Poisson equations.
In the Hernquist method \citep{Hernquist1993}, the velocity distribution
of the stellar and dark matter components are assumed to be \textcolor{black}{Gaussian}
with a dispersion found by solving the Jeans equations. Since solutions
to the Jeans equations are not true solutions to the collisionless Boltzmann
equation, these models typically relax to a state different
from the one specified in the ICs \citep{Kazantzidis2004}.
Other methods \textcolor{black}{include \textsc{mkgalaxy} \citep{McMillan2007}, which uses a guided relaxation
algorithm, \textsc{GalIC} \citep{Yurin2014}, whose algorithm has elements of the
Schwarzchild method \citep{Schwarzchild1979} of building orbit
libraries, and the iterative method presented in
\citet{Rodionov2009,Rodionov2011}}.

In contrast with the Hernquist method, the \textsc{GalactICS} DFs are based
on the Jeans theorem, which states that functions of the integrals of
motion are equilibrium solutions to the collisionless Boltzmann equation (CBE).
In particular, the DFs for the bulge and halo are assumed to be functions
of the energy, which are constructed to yield the \textcolor{black}{target}
density profiles via the Eddington inversion formula \citet{BT2008}.
The DF for the \textcolor{black}{disc} is a function of the energy, angular momentum about
the symmetry axis, and the vertical energy. Since the latter is only approximately
conserved, the DFs are not exact solutions to the CBE. However, for 
thin \textcolor{black}{discs}, the approximation is excellent, and the model ICs for even
thick \textcolor{black}{discs} are are relatively stable. Armed with DFs in terms
of the integrals of motion, \textsc{GalactICS} solves the Poisson equation
through an iterative algorithm, \textcolor{black}{which meshes well with the \citet{Wang2010} prescription for building an equilibrium gas \textcolor{black}{disc}}.
The code has been used to study
bar and spiral structure formation in the Milky Way (MW) \citep{Widrow2008, Fujii2019},
bending waves in \textcolor{black}{discs} \citep{Chequers2017}, how radial 
dispersions and the Galactic bar influence stellar populations
\citep{Debattista2017}, merging dwarf galaxies \citep{Lokas2014}, and more.

Recently, a new method based on angle-action variables called
\textsc{AGAMA} \citep{Vasiliev2019}
has become available. It also involves
an iterative scheme to solve Poisson's equation with the additional
step that for each iteration, one must construct that action
variables in the new potential. One advantage of the method 
over \textsc{GalactICS} is that it avoids the approximate "third integral", $E_z$
and therefore should be able to do a better job of modelling warm \textcolor{black}{discs}.
On the other hand, the connection between angle-action variables and
the usual phase space coordinates of less transparent. Thus, it can be 
more difficult to build models with specific structural properties, such
as a stellar \textcolor{black}{disc} with constant scale height (see, for example, \citep{Chequers2018}).
In principle, it should be straight forward to add a gas \textcolor{black}{disc} to an \textsc{AGAMA} model using a method similar to the one described in this paper.

\textcolor{black}{Yet another approach can be found in \citet{Rodionov2011} who introduced an alternative iterative approach that allows one to build non-equilibrium galaxy models, such as
\textcolor{black}{discs} embedded in triaxial halos. In their scheme, the system is evolved on a short
time scale via the standard dynamical equations. Structure properties or parameters of the system,
such as the surface density, are then reset and the system is again allowed to evolve.
Ultimately, the sequence produces a system that is close to equilibrium.}

The outline of the paper is as follows:
In Section \ref{Sec:GalactICS}
We present the details of this new version 
of \textsc{GalactICS}. We then examine two gas-halo models in
Section \ref{Sec:GasModels}. The models are similar to the ones consideblack
in \citet{Wang2010} except in our simulations, the halo is 'live'.
In Section \ref{Sec:MWModel}, we
construct a 5-component MW model based on the best-fit model from 
\citet{McMillan2017}. We conclude with a discussion of our results in Section \ref{Sec:Con}.

\section{GalactICS Models} \label{Sec:GalactICS}

The heart of \textsc{GalactICS} is the construction of a
DF for each of the components that together yield an axisymmetric model
with the desired structural and kinematic properties. \textcolor{black}{The total DF is given by}
\begin{equation}\label{Eq:DF}
\begin{split}
 f(E,L_{z},E_{z})=& f_{b}(E)+f_{h}(E)
  +f_{d,1}(E,L_{z},E_{z})\\
 &+f_{d,2}(E,L_{z},E_{z})+f_{g}(E,L_{z},E_{z})~,
 \end{split}
\end{equation}
where $E$ is the energy, 
$L_{z}$ is the angular momentum about the symmetry axis of the system, and $E_{z}$ is the 
energy of the vertical motions in the \textcolor{black}{discs}.
For a time-independent and axisymmetric system, $E$ and $L_{z}$ are 
conserved while $E_{z}$ is 
only approximately conserved for \textcolor{black}{disc} particles on nearly circular orbits.

\textcolor{black}{The density $\rho$ is determined by the integral of $f$ over all velocities
and since $f$ is a function of $E$, which depends on $\Phi$, Poisson's equation}
is an implicit function of the $\Phi$:
\begin{equation}\label{Eq:GenPot}
 \nabla^{2}\Phi=4\pi \rho(R,z,\Phi) = 4\pi \int d^3v \,f\left (E,\,L_z,\,E_z\right )~.
\end{equation}
Note that
here and throughout, we set Newton's constant $G=1$.
\textsc{GalactICS} numerically solves 
Eq. \ref{Eq:GenPot} using
an iterative approach to obtain a self-consistent
density-potential pair.  First a target density
profile is selected for each component. 
(The gas \textcolor{black}{disc} is treated somewhat differently. See Sec. \ref{subsec:Gasdisc}). The corresponding
 potential is found by solving Poisson's equation \textcolor{black}{using an expansion in Legendre
 polynomials. The result is used, in turn, to calculate a new density 
and the process is repeated until the density-potential pair converges}.

 In detail, most of the algorithm is identical to that 
 given in \citet{Widrow2008}. By design, the bulge has
 a S\'{e}rsic density profile, the \textcolor{black}{disc} components are
truncated exponential-sech$^2$ profiles, and the halo is a 
 truncated double-power law.  As with 
 \citet{Widrow2008}, the bulge and halo use 
 the Abel integral transformation \citep{BT2008}
 to get their DF, 
\begin{equation}\label{Eq:SpheblackF}
 f_{i}(\epsilon)=\frac{1}{\sqrt{8\pi^{2}}}
 \int_{E}^{0}\frac{d^{2}\tilde{\rho}_{i}}{d\Phi^{2}}
 \frac{d\Phi}{\sqrt{\Phi-E}}~,
\end{equation}
where $i$ is either the bulge or halo.
In the presence of 
any of the \textcolor{black}{disc} components, this equation is solved using 
a spherical approximation of the \textcolor{black}{disc} potential for
$\Phi_{tot}$.  It is important to note that this approximation
is not used for the calculation of $\Phi(R,z)$ or $\rho(R,z)$, but is
only used for Eq. \ref{Eq:SpheblackF}.  It is also worth noting that this
method of evaluating Eq. \ref{Eq:SpheblackF} causes a degree of 
flattening, but the spheroid components remain isotropic and 
axisymmetric.

\subsection{Stellar \textcolor{black}{discs}}\label{subsec:disc}

The flattened nature of the stellar \textcolor{black}{discs} demands a
different approach than the simple solutions used 
for the bulge and halo components. \textsc{GalactICS} uses the 
DF presented in \citet{Kuijken1995}, which itself
is based on \citet{Shu1969} and \citet{Binney1987}.

The potential for this flattened system is calculated 
using a combination of spherical harmonics and 
an analytic 'fake' density-potential pair \citep{Kuijken1995}.
This pair, $(\rho_{fd},\Phi_{fd})$, has the property that
$\rho_{d}=\rho_{fd}+\rho_{r}$ and 
$\Phi_{d}=\Phi_{fd}+\Phi_{r}$ where $\rho_{r}$ and $\Phi_{r}$ 
are residuals.  The 'fake' components are designed to account
for the higher order moments of the total potential, while the 
lower order moments are calculated by solving Poisson's
equation for the residuals using a small number 
of $l$ moments.  The two are then summed together to give
the total potential of the \textcolor{black}{disc} given some density.  In 
practice, we find that $l_{max}=10$ is sufficient for 
most models.

The \textsc{GalactICS} DFs for the \textcolor{black}{disc} components are constructed
to yield a vertical structure that is approximately isothermal.
Thus, the vertical velocity dispersion is related to the thickness of the
\textcolor{black}{disc} and the surface density. The 
DF is designed to yield a scale height that is approximately constant
across the \textcolor{black}{disc} and a surface density that is approximately exponential
with a scale radius $R_d$. Thus, the vertical dispersion profile is
given by 
\begin{equation}
\sigma_z^2 \simeq \sigma_{z,0}^2 e^{-R/R_d}
\end{equation}
The radial dispersion profile appears as a free function in the \textcolor{black}{disc}
DF. Here, we assume
\begin{equation}
 \sigma^{2}_{R}(R)\simeq\sigma^{2}e^{(-R/R_{d})}~.
\end{equation}
\textcolor{black}{This is based on the observations of \citet{Bottema1993}.}
The tangential dispersion is found through the epicycle approximation:
\begin{equation}
 \sigma_{\phi}(R)\simeq\sigma_{R}(R)\frac{\kappa}{2\omega}~,
\end{equation}
where $\omega$ is the angular frequency and $\kappa$ is the epicyclic frequency.
Note that in the actual DF, $\sigma_R$, $\kappa$, and $\omega$ are written
as functions of the guiding radius $R_c$, which is a function of $L_z$ and hence
an integral of motion. The approximations in the above equations reflect the
fact that $R_c$ is only approximately equal to $R$.

\subsection{Gas \textcolor{black}{disc}}\label{subsec:Gasdisc}

\citet{Wang2010} introduced two methods for generating isothermal
equilibrium gas \textcolor{black}{discs} in a general galactic potential. We have adapted their 
'potential' method. Since the \textcolor{black}{disc} is isothermal, the 
scale height increases as a function of the radius. 
We assume a target 
exponential surface density for gaseous component, which
causes the space density in the
midplane to be a decreasing function of radius.

Following \citet{Wang2010}, we assume that the 
density of the gas \textcolor{black}{disc} is given by
\begin{equation}
 \rho_{g}(R,z)=\rho_{0}(R)\textrm{exp}\left(-\frac{\Phi_{z}(R,z)}{(\gamma-1)\epsilon}\right)~,
\end{equation}
where $\rho_{0}(R)$ is the mid-plane density, $\gamma$ is the adiabatic index, $\epsilon$ is
the specific internal energy, and $\Phi_{z}(R,z)=\Phi(R,z)-\Phi(R,0)$.
The specific internal energy depends on the temperature through the equation of state
\begin{equation}
 \epsilon=\frac{1}{\gamma-1}\frac{k_{B}T}{\mu m_{p}}~,
\end{equation}
where $k_{B}$ is the Boltzmann constant, 
$\mu$ is the atomic weight of the gas, and 
$m_{p}$ is the mass of the proton. For simplicity, that gas is assumed to be 
hydrogen and adiabatic so $\mu=1$, $\gamma=5/3$, and $\epsilon$ only depends 
on the gas temperature. It is convenient to re-parameterize
the gas temperature as\textcolor{black}{
\begin{equation}
 c_{0}=\frac{k_{B}T}{\mu m_{p}}~,
\end{equation}
which is similar to the specific internal energy and related 
to the sound speed}. Combining the target exponential surface density, 
\begin{equation}\label{Eq:sigR}
 \Sigma(R)=\Sigma_{0}e^{-R/R_{g}}=\int_{-\infty}^{\infty}\rho_{g}(R,z)dz~,
\end{equation}
with $\rho_{g}(R,z)$ gives 
\begin{equation}
 \rho_{0}(R)=\frac{\Sigma_{0}e^{-R/R_{g}}}
 {\int_{-\infty}^{\infty} e^{[-\Phi_{z}/c_{0}]}dz}~.
\end{equation}
The result is then included in Poisson's equation
(Eq.\,\ref{Eq:GenPot}). The algorithm for solving this equation
proceeds as follows: First an initial surface density profile is used to estimate
the gas \textcolor{black}{disc} potential. That potential is then used to calculate the
scale height profile and total density, which are in turn 
used to get a new surface density and potential. 
The whole process is repeated until
the system converges. Note that the surface density of the final system
may deviate slightly from the target exponential surface density.
As with the stellar \textcolor{black}{disc}, we use
an analytic fake density-potential pair when solving 
Poisson's equation.

The gas \textcolor{black}{disc} density-potential pair is constructed at the 
same time as the total density-potential pair.  The full \textsc{GalactICS}
algorithm for obtaining a self-consistent density potential pair is 
then given as follows:
\begin{enumerate}
\item Define the target profiles for all components.
\item Estimate the total potential using Poisson's equation.
\item Calculate new densities for the collisionless components
using their DFs.
\item Calculate the gas density using the current scale height
profile and total potential.
\item Calculate new gas surface density and scale height profiles.
\item Repeat steps ii-v to achieve convergence.
\end{enumerate}

We use a greatly simplified velocity structure for the gas \textcolor{black}{disc}
compared to the stellar \textcolor{black}{discs}.  The gas temperature is 
treated as a kinetic temperature that accounts for all turbulent
motions.  Therefore, the particles are initialized on 
\textcolor{black}{purely rotational orbits
where the speed} is
\begin{equation}\label{Eq:GasVel}
 V^{2}(R,z)=R\left. \frac{\partial \Phi}{\partial R}\right|_{z=0}
 +c_{0}\left. \frac{\partial \ln \rho }{\partial \ln R} \right|_{z=0}~,
\end{equation}
with no random motions.  \textcolor{black}{Near the center it is
possible for the gas pressure term to set $V^{2}<0$.
While this condition has not occurred in any of the 
subsequent simulations, \textsc{GalactICS} currently 
sets $V^{2}=0$ if the gas pressure term yields a negative
result.
In such cases we set $V^{2}=0$.  An alternate method is 
to use the softening procedure outlined in \citet{Hernquist1993}.
At this point we have not yet implemented and tested this in the 
context of \textsc{GalactICS}.}  In addition, \textsc{GalactICS} sets $v_{z}=0$ for 
all gas particles as the vertical gravitational
force is balanced against the gas pressure.

While this section has focused on the generation of a gas 
\textcolor{black}{disc} within the \textsc{GalactICS} methodology, it is worth noting that
this same algorithm can be included in the \textsc{AGAMA} method with
little in the way of modifications
since \textsc{AGAMA} also uses
an iterative approach to calculate the total potential and
density.\textcolor{black}{
However, since \textsc{GalIC}, 
and the Hernquist method do 
not include an iterative adjustment of the total potential,
it would be difficult to modify those algorithms
to utilize this \textcolor{black}{particular} method of constructing a gas \textcolor{black}{disc}.  It would
also be difficult to implement this method of 
gas disc construction into \textsc{mkgalaxy} due to
differences in how the iterative calculation of the potential
is performed.  Nonetheless,
gas discs could be generated in all these algorithms using
alternate methods than the process utilized here.
}
\section{Gas Only Models} \label{Sec:GasModels}

In this section we describe testbed simulations that involve
a gas \textcolor{black}{disc} and dark halo. The ICs roughly correspond to the Gas0
and Gas4 models of \citet{Wang2010}. They have the same structural
properties for the two components and differ only in the gas temperature.
In particular, the dark halo has an NFW profile with scale length
$r_h = 17.8\,{\rm kpc}$ and velocity scale parameter (in \textsc{GalactICS} parameterization)
$\sigma_h = 389\,{\rm km\,s}^{-1}$. This velocity scale parameter implies
an NFW density parameter of $\rho_0 = \sigma_h^2\left (2\pi r_h\right )^{-2} \simeq 1.77\times 10^7\,M_\odot\,{\rm kpc}^{-3}$.
The gas \textcolor{black}{disc} has a mass of $M_d = 10^{10}\,\Msol$
and an exponential scale length $R_g = 3.5$ kpc. The resultant rotation
curve rises slowly to about $180\,{\rm km\,s}^{-1}$ at a radius of $30$ kpc.
The gas contribution to the rotational speed is always subdominant indicating that
the \textcolor{black}{disc} will be stable to global perturbations. Following \citet{Wang2010}
we set the gas temperature in the GalactICS-Gas0 model
to $T=4\times10^{4}$ K and the temperature in the
GalactICS-Gas4 model to $T=8\times10^{3}$ K. With these values,
we expect the GalactICS-Gas0 model to be stable to local perturbations
and the colder GalactICS-Gas4 model to be unstable (see Figure 4 of \citet{Wang2010}).

The DFs for the models are sampled with
$1$M gas particles and $5$M halo particles
and evolved for 1.6 Gyr using the Gadget 2 N-body code \citep{Springel2005}.
\textcolor{black}{We use an adaptive softening length with an initial length 
set to 0.005 kpc and a maximal softening lengths of 0.5 kpc.
This is smaller than the recommended softening 
lengths from \citet{Rodionov2005}.  However, the gas \textcolor{black}{discs} for both models are
quite thin in the central kpc.  As such, we feel
that this small scale length is justified.
In the simulation, we use an adaptive time step with a maximal 
step of 0.01 Gyr (which also matches the snapshot frequency).  
The Courant factor in the \textsc{Gadget 2} simulation is set 
equal to 0.25.}

 We stress that in our simulations, the halo
is 'live' whereas the halo in the 
\citet{Wang2010} simulations is treated as
a static halo potential. Moreover, \citet{Wang2010} ran their simulations
with the adaptive-mesh-refinement code RAMSES \citep{Teyssier2002}
whereas \textsc{Gadget 2} models the gas using smooth particle
hydrodynamics.

\subsection{The GalactICS-Gas0 model} \label{subsec:Gas0}

\begin{figure}
\centering
    \includegraphics[width=80mm]{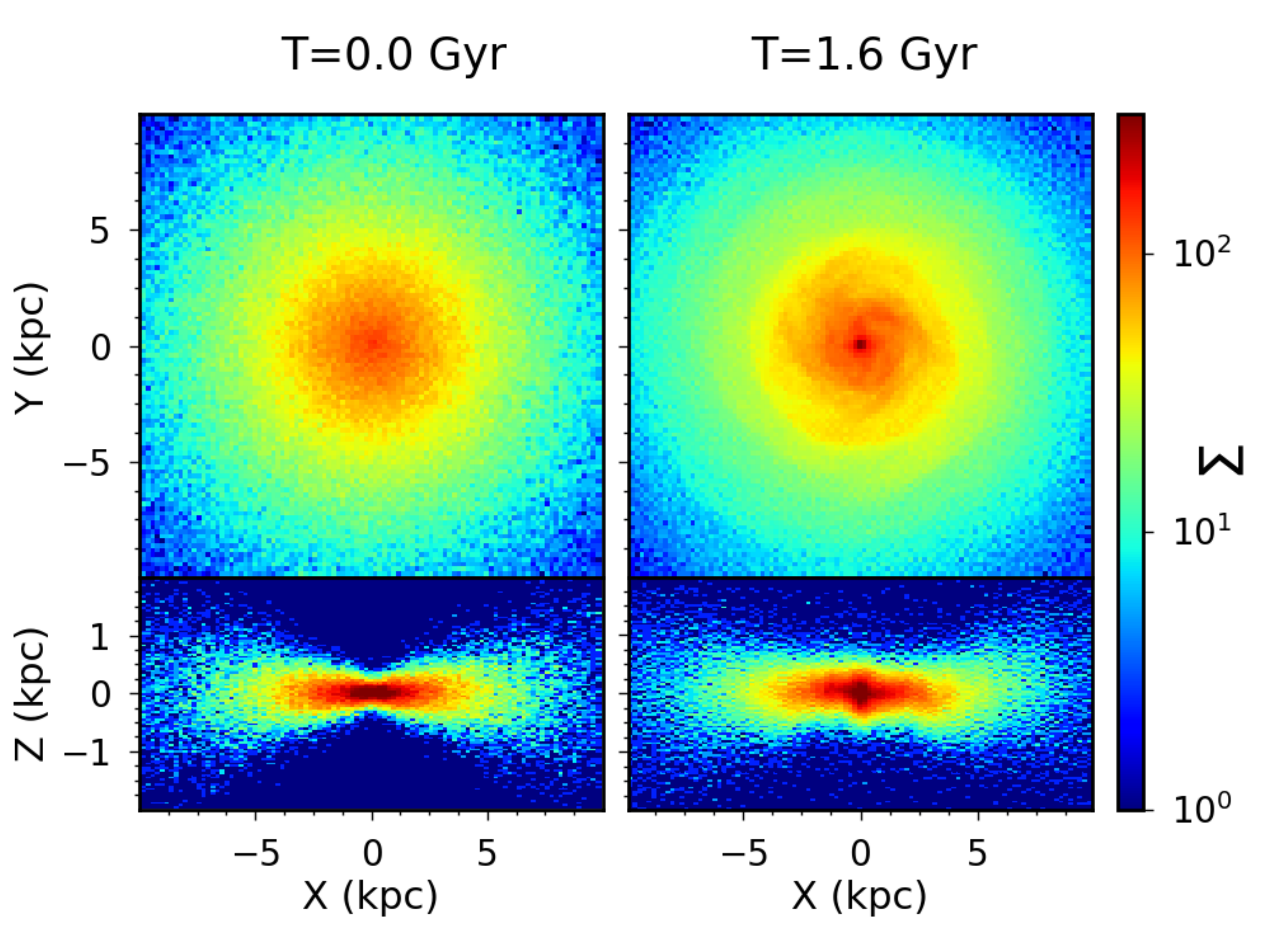} 
\caption{The surface density (upper panel) and cross-sectional density (lower panel) of the GalactICS-Gas0 model 
gas \textcolor{black}{disc} at $T=0$ Gyr (left) and $T=1.6$ Gyr (right).  The units are $M_{\odot}/pc^{2}$
and $M_{\odot}/pc^{3}$ respectively.}
  \label{Fig:Gas0Img}
\end{figure}

Figure \ref{Fig:Gas0Img} shows the initial and final surface
and cross-sectional densities for the GalactICS-Gas0 model. It is 
clear that there has been some evolution of the system. In particular, the central
surface density of the \textcolor{black}{disc} has increased (upper panels), as has its scale height
(lower panels).  

\begin{figure}
\centering
    \includegraphics[width=80mm]{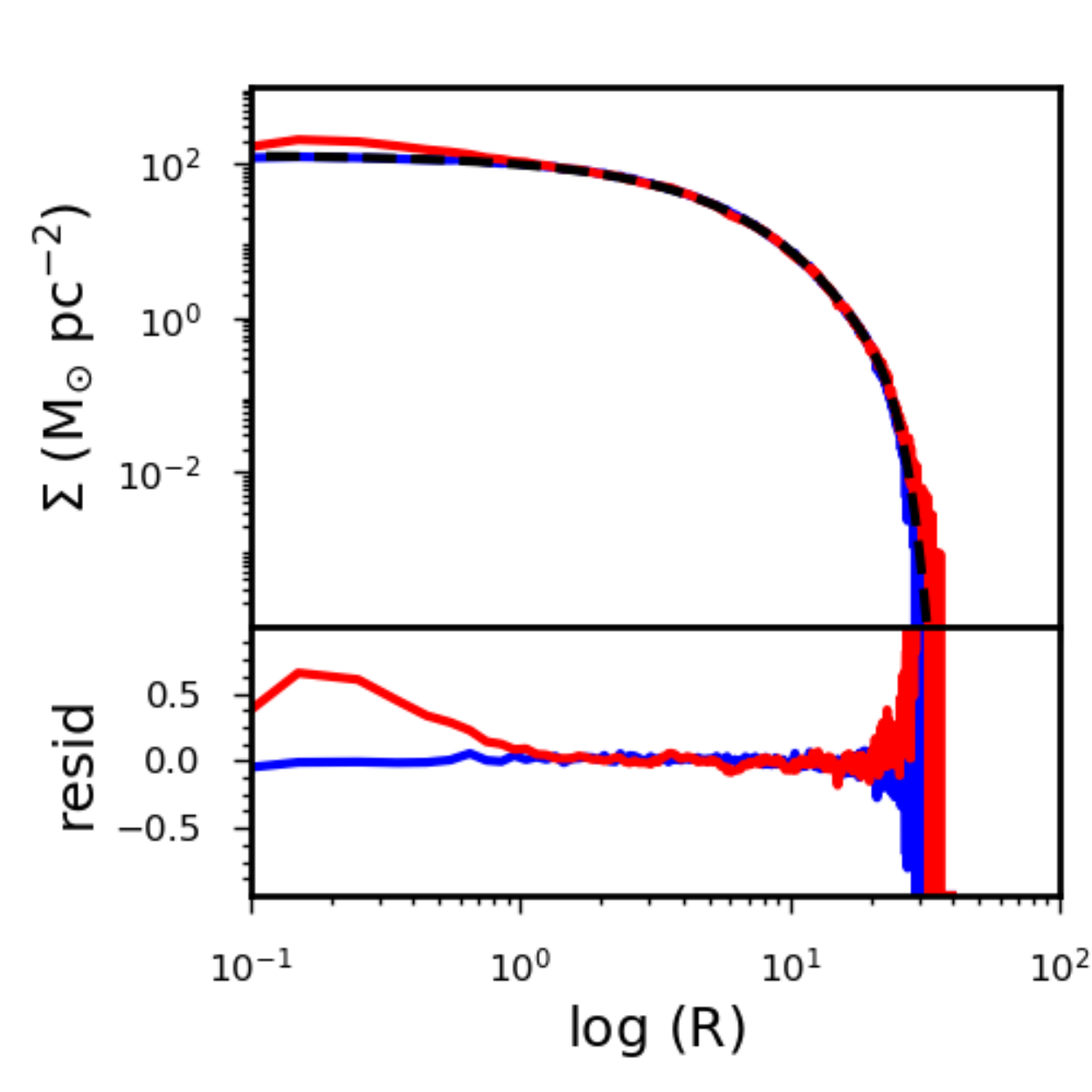} 
\caption{The azimuthally averaged surface density of the GalactICS-Gas0 gas \textcolor{black}{disc} as a function of radius (top) and the density residuals (bottom).  The dashed
black curve is the initial analytic profile and the solid blue and red lines are the surface densities
at $T=0$ and $1.6$ Gyr respectively.  The distance units are kpc.}
  \label{Fig:Gas0_SD}
\end{figure}

\begin{figure}
\centering
    \includegraphics[width=80mm]{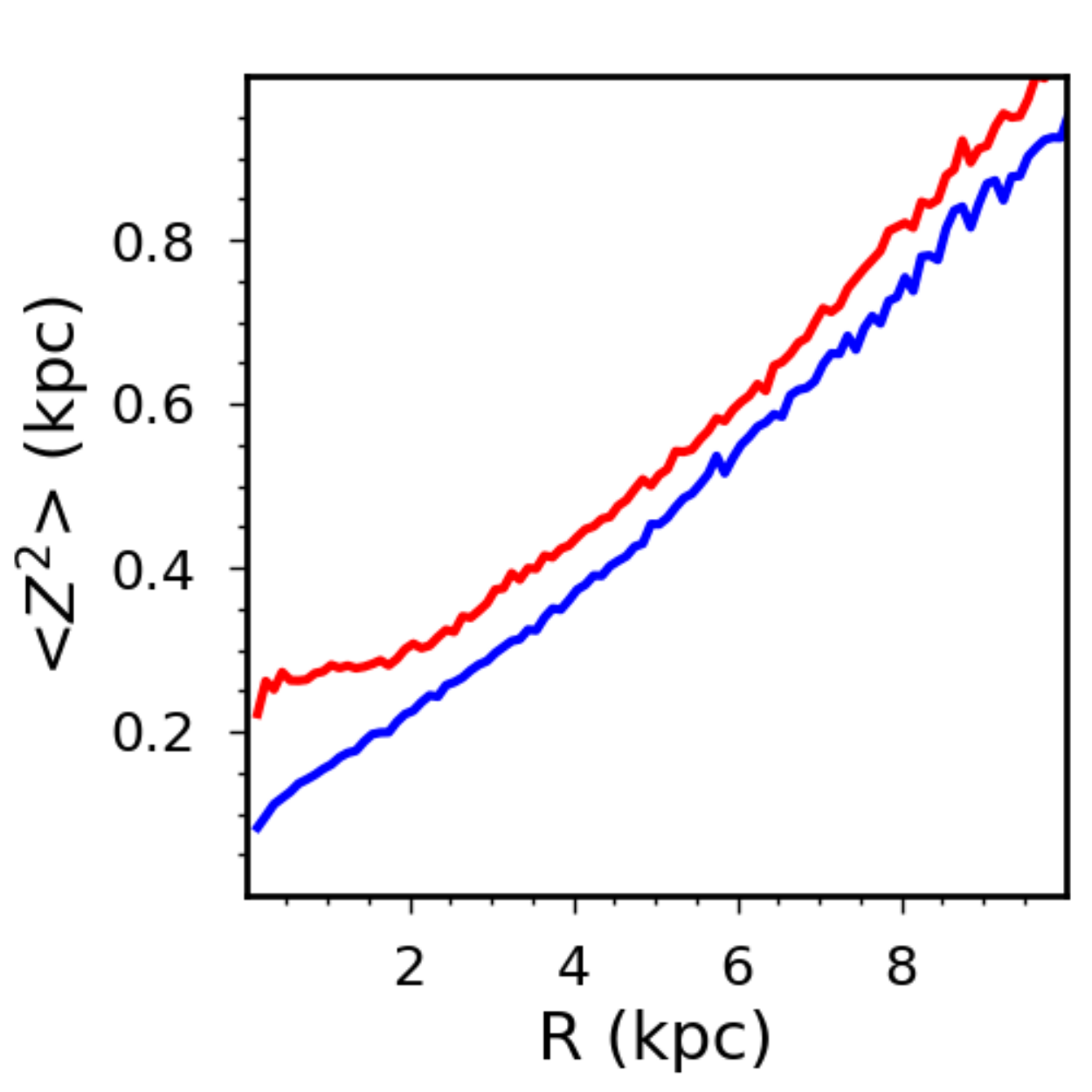} 
\caption{The average gas \textcolor{black}{disc} thickness of the GalactICS-Gas0 
model as a function of 
radius for $T=0$ Gyr (blue) and $T=1.6$ Gyr (red).}
  \label{Fig:Gas0Vert}
\end{figure}

The changes in the gas surface density and thickness 
are highlighted in Figs. \ref{Fig:Gas0_SD} and \ref{Fig:Gas0Vert}
respectively. In Fig. \ref{Fig:Gas0_SD} we see that the surface 
density has increased significantly in the central regions. In addition,
the sharp cut-off at the edge of the \textcolor{black}{disc} has been smoothed out.
On the other hand, at intermediate radii ($1<R<15\,{\rm kpc}$), the evolved surface
density is consistent with the initial one.
The \textcolor{black}{disc} thickness increases at all radii,
with the largest increase occurring in the central $0.5$ kpc.

\begin{figure}
\centering
    \includegraphics[width=80mm]{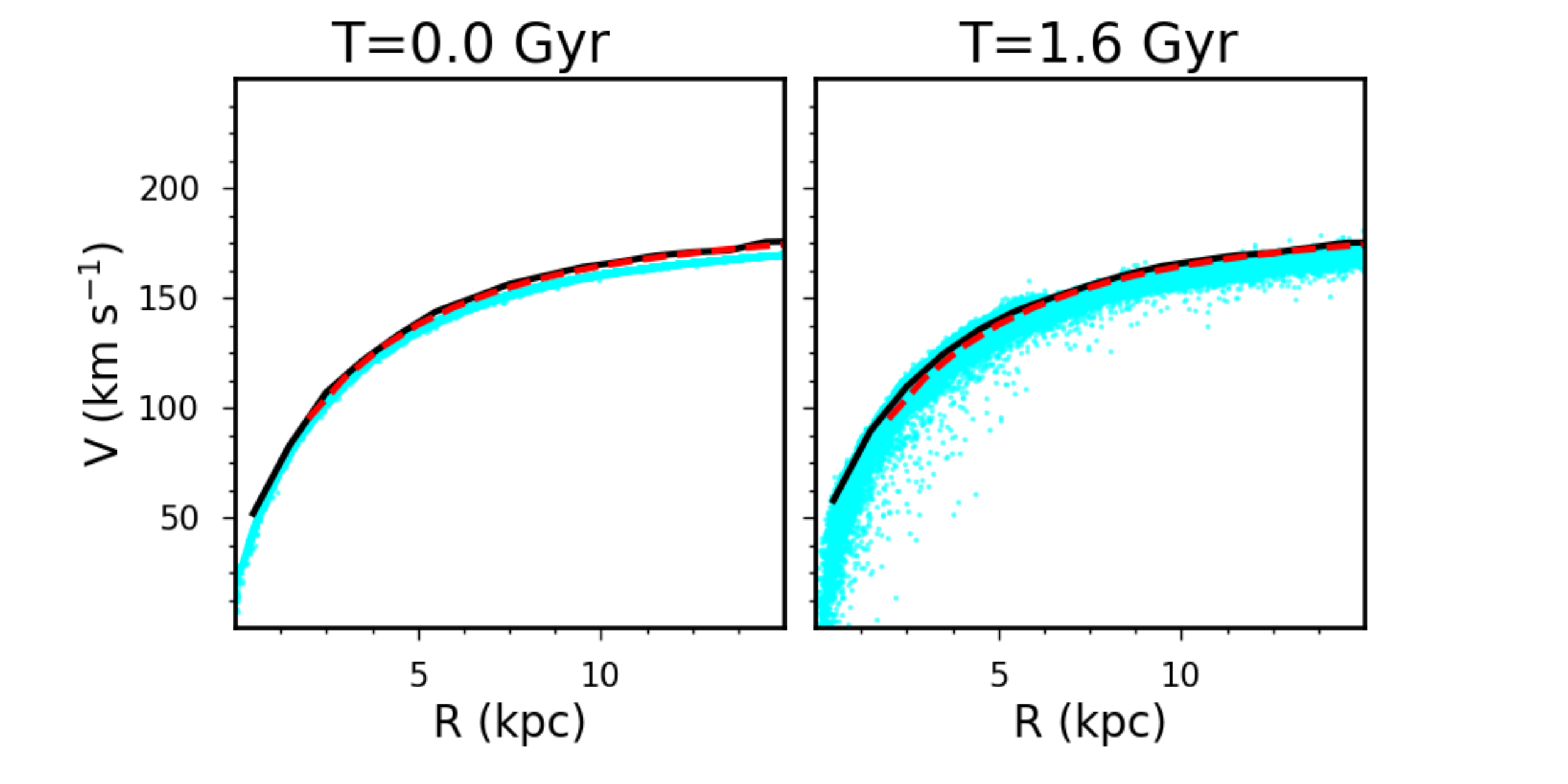} 
\caption{The rotation curve of the GalactICS-Gas0 model at $T=0$ (left) and 1.6 (right) Gyr.  The dashed red curve is the initial
analytic velocities, the solid black line is the expected velocity from the azimuthally averaged radial force, 
the cyan points at the tangential velocities of particles.}
  \label{Fig:Gas0_RC}
\end{figure}

Further clues as to the evolution of the system can be found
in Fig. \ref{Fig:Gas0_RC}, which shows that the 
azimuthally averaged rotation curve \textcolor{black}{(calculated
from the gravitational potential)} as well as 
a scatter plot of gas
particle azimuthal velocities. \textcolor{black}{Although the gas \textcolor{black}{disc}
is initialized with purely 
rotational motions, it is clear from the figure
that the gas particles gain random azimuthal
(and presumably radial and vertical)
motions.}
These random motions represent a transfer of rotational
energy to the random kinetic energy of the gas particles.
It is therefore not surprising that the \textcolor{black}{disc} 
"puffs up" a bit and also compresses slightly in the radial
direction, thereby increasing the central surface density.
Note that the \textcolor{black}{gravitational} rotation curve stays 
very nearly constant 
over the duration of the simulation.

\begin{figure}
\centering
    \includegraphics[width=80mm]{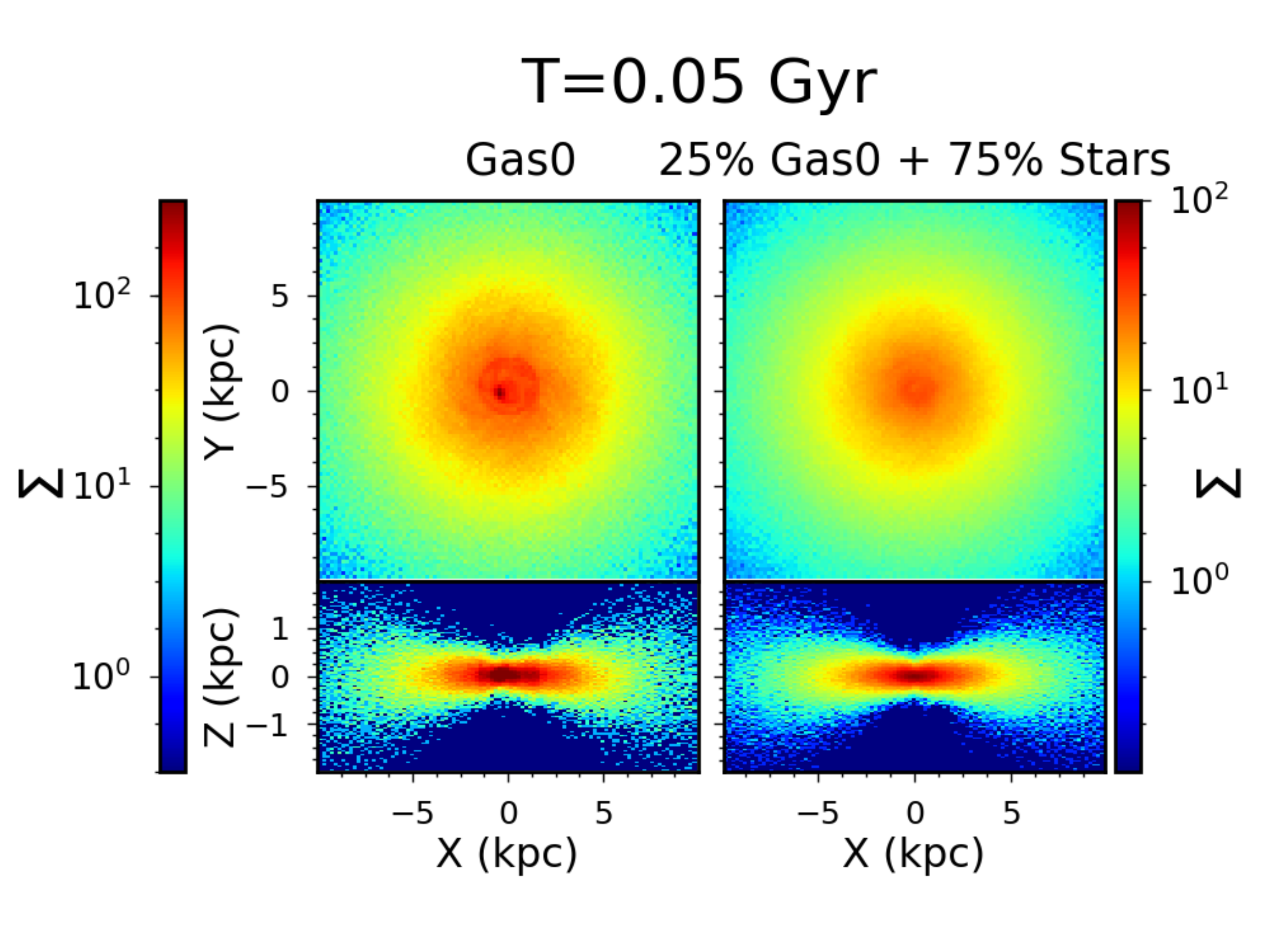} 
\caption{The surface density (upper panels) and cross-sectional density (lower panels) of the 
gas \textcolor{black}{disc} for the GalactICS-Gas0 model and Quarter-Gas0 model 
at $T=0.05$ Gyr.  The differing scales for the left and right panels is to highlight the 
over-density transient present in the Gas0 model.  The units of surface density and cross-sectional
density are $M_{\odot}/pc^{2}$
$M_{\odot}/pc^{3}$ respectively.}
  \label{Fig:QuarterGas0-Gas0Comp}
\end{figure}

To investigate the transients in this simulation we show, in
Fig. \ref{Fig:QuarterGas0-Gas0Comp},
the face-on and edge-on projections of the density after just
50 Myr. A ring of particles corresponding to a axisymmetric density
wave, is clearly visible.

Apart from the scattering of \textcolor{black}{disc} particles, a second possible
cause of the transients may be in the use of the 'fake' gas \textcolor{black}{disc}
density-potential pair. The gas \textcolor{black}{disc} is extremely thin near its
centre with a scale height of less than 100 pc. Moreover, the 
thickness changes rapidly with radius there. Thus, it may be difficult
for our fake-\textcolor{black}{disc}/Legendre polynomial expansion scheme to full 
capture the potential. 
Small errors in the potential solver
can lead to an incorrect scale height and 
an over/under pressure, thereby causing a transient wave, as is seen
in Fig. \ref{Fig:QuarterGas0-Gas0Comp}.

To be clear the approximation that causes the transient only fails
in regions where the value of $\Phi_{z}$ is dominated by the 
gas \textcolor{black}{disc} contribution.  At moderate radii the halo dominates
$\Phi_{z}$ and the small errors from the gas potential
approximation are inconsequential.
To illustrate this point, we generated
a Gas0-stars model where the gas \textcolor{black}{disc} is replaced by a 
two-component \textcolor{black}{disc} of gas and stars with a mass ratio of $1:3$.
The velocity dispersion for the stars is assumed to be exponential
dispersion as in Eq. \ref{Eq:sigR} with a central dispersion of $\sigma_{r,0}=100 \kms$.
The right panel of Fig. \ref{Fig:QuarterGas0-Gas0Comp} shows the 
gas \textcolor{black}{disc} of this system at 50 Myr.  In contrast with the GalactICS-Gas0 model,
there is no transient wave even though the system has the same 
surface density profile. It is worth noting that 
the thickness of the stellar \textcolor{black}{disc} also thickens the gas \textcolor{black}{disc} in the 
central regions while decreasing the degree of flaring in the 
outer regions.

\subsection{The GalactICS-Gas4 Model}\label{subsec:Gas4}

We next consider the GalactICS-Gas4
model, which is unstable to local perturbations. In the \citet{Wang2010}
realization of this model, the gas \textcolor{black}{disc} rapidly 
fragments (see the lower left panel of their Figure 5).
\textcolor{black}{Our realization of this model is also unstable, but it evolves
slightly differently.}

\textcolor{black}{
Fig. \ref{Fig:Gas4Img} shows our gas \textcolor{black}{disc} at 50 Myr and
1 Gyr.  The initially very thin \textcolor{black}{disc} thickens significantly
by 1 Gyr.  At 50 Myr, the \textcolor{black}{disc} has begun fragmenting into 
a strong spiral structure.  This fragmentation is less 
symmetric than that seen in the 
\citet{Wang2010} model due to both the random sampling 
procedure of \textsc{GalactICS} and the live halo.  By
1 Gyr this instability has evolved in such a way to
produce thick central region with poorly defined
spiral arms.}

\textcolor{black}{
\begin{figure}
\centering
    \includegraphics[width=80mm]{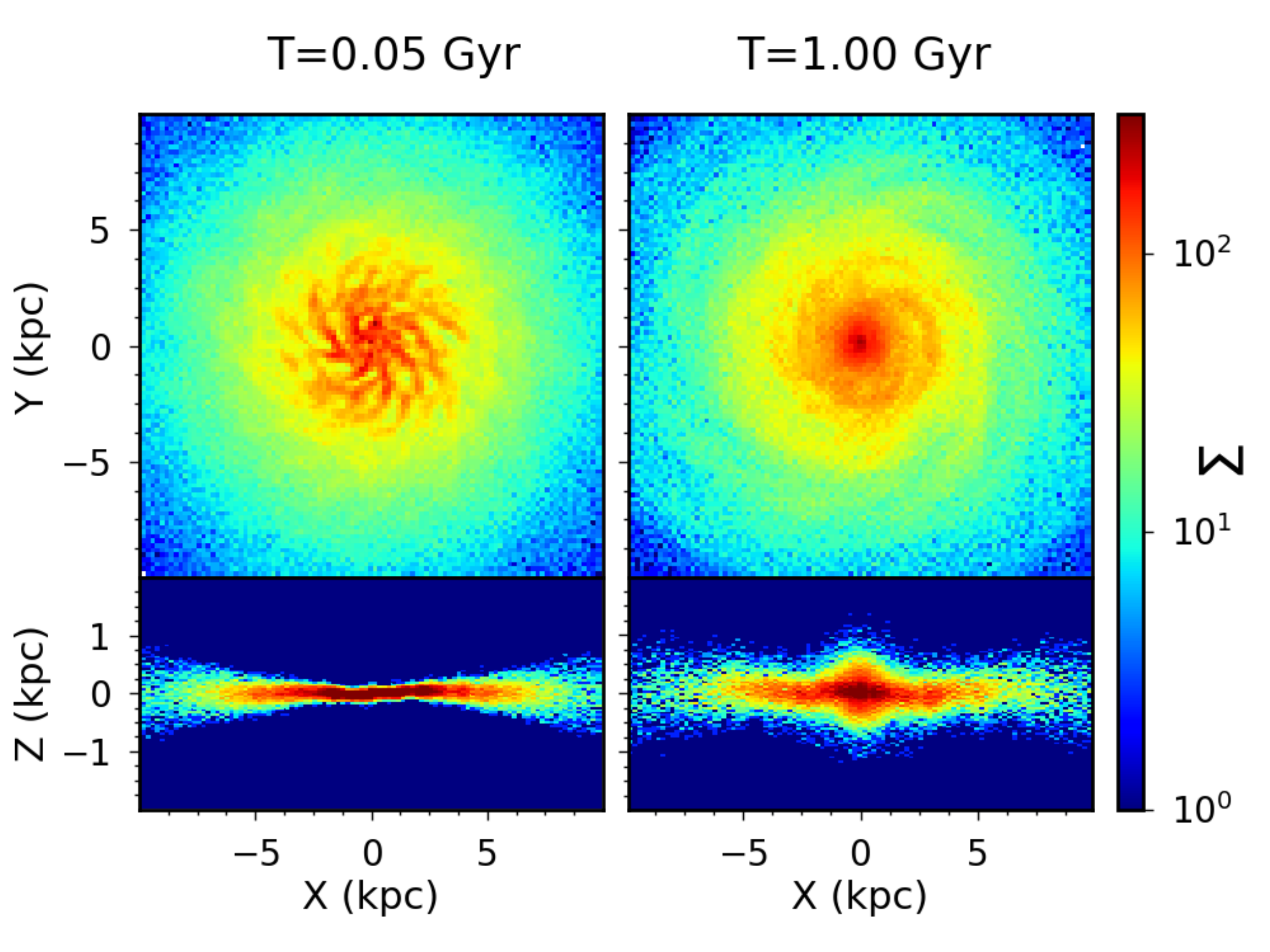} 
\caption{The surface density (top) and cross-sectional density (bottom)
of the GalactICS-Gas4 gas \textcolor{black}{disc} at $T=0.05$ and $T=1.0$ Gyr.
The surface density and cross-sectional
density units are $M_{\odot}/pc^{2}$
$M_{\odot}/pc^{3}$ respectively.}
  \label{Fig:Gas4Img}
\end{figure}
}
\textcolor{black}{
Figure \ref{Fig:Gas4Vert} shows the initial \textcolor{black}{disc} thickness and
the thickeness at $T=1.0$ Gyr.  Comparing this to Fig. 
\ref{Fig:Gas0Vert} it is clear that the Gas4 gas \textcolor{black}{disc} is initially
about half the thickness of the GalactICS-Gas0 \textcolor{black}{disc}.  
The height in the central region increases to about the same
as the Gas0 model in the final timesteps, but rather than staying
at some constant thickness before flaring in the outer radii, the
Gas4 thickness decreases to a minimum near 3 kpc before 
flaring to the outer radii.  
}

\textcolor{black}{
\begin{figure}
\centering
    \includegraphics[width=80mm]{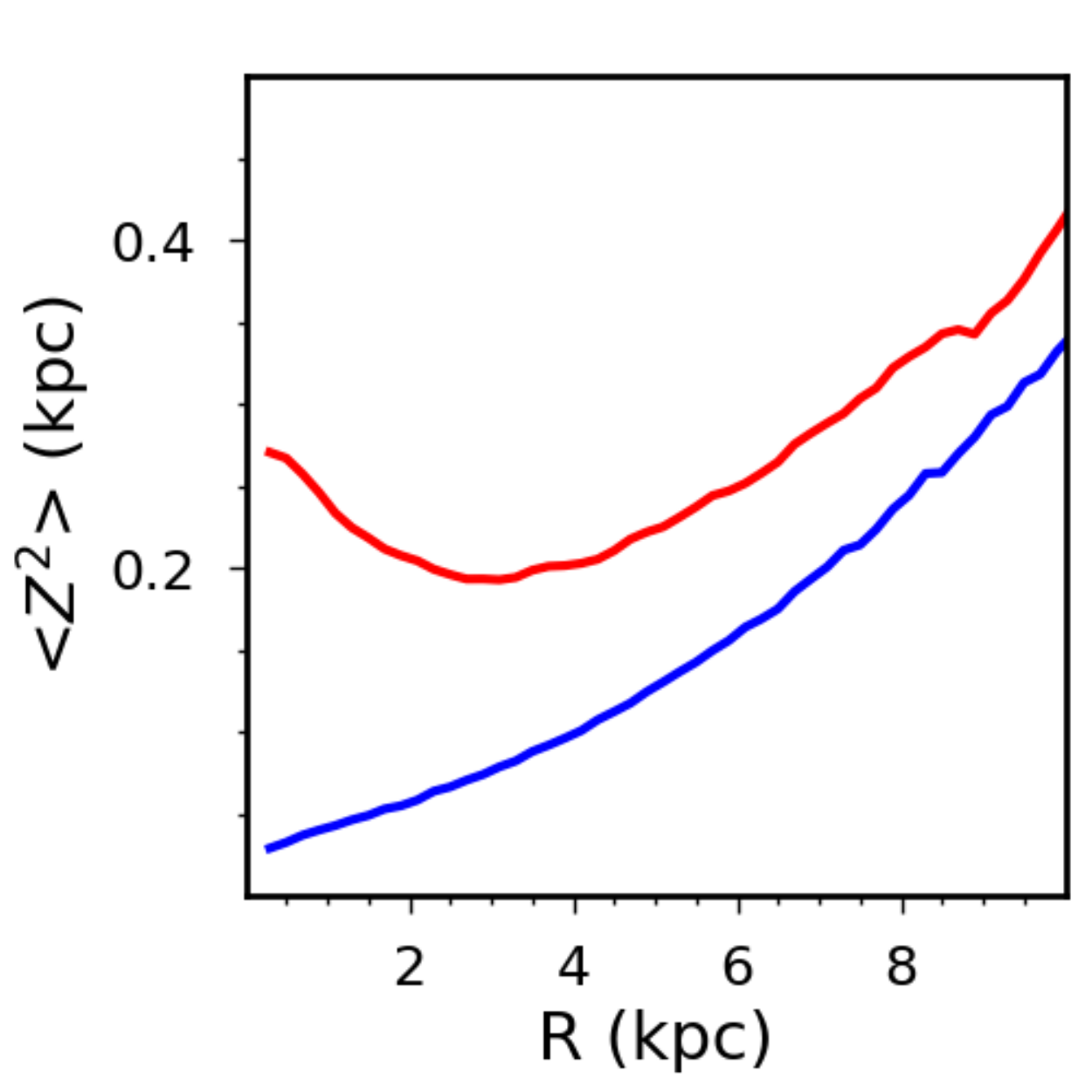} 
\caption{The average gas \textcolor{black}{disc} thickness of the GalactICS-Gas04
model as a function of 
radius for $T=0$ Gyr (blue) and $T=1.0$ Gyr (red).}
  \label{Fig:Gas4Vert}
\end{figure}
}

\textcolor{black}{
Figure \ref{Fig:Gas4_SD} shows the evolution of the
azimuthally averaged surface density profile.
The changes to the averaged profile are similar in 
scale and origin 
to the changes that occur for the GalactICS-Gas0 
model.  
}

\textcolor{black}{
It is important to note that both the 
GalactICS-Ga0 and Gas4 models are fairly different
from real galaxies.  Nonetheless,
these two models demonstrate the ability of 
 \textsc{GalactICS} to generate both stable and
 unstable models.  These models are relatively 
 straightforward, consisting of only a halo and 
 a thin (or very thin) gas disc.
}

\textcolor{black}{
\begin{figure}
\centering
    \includegraphics[width=80mm]{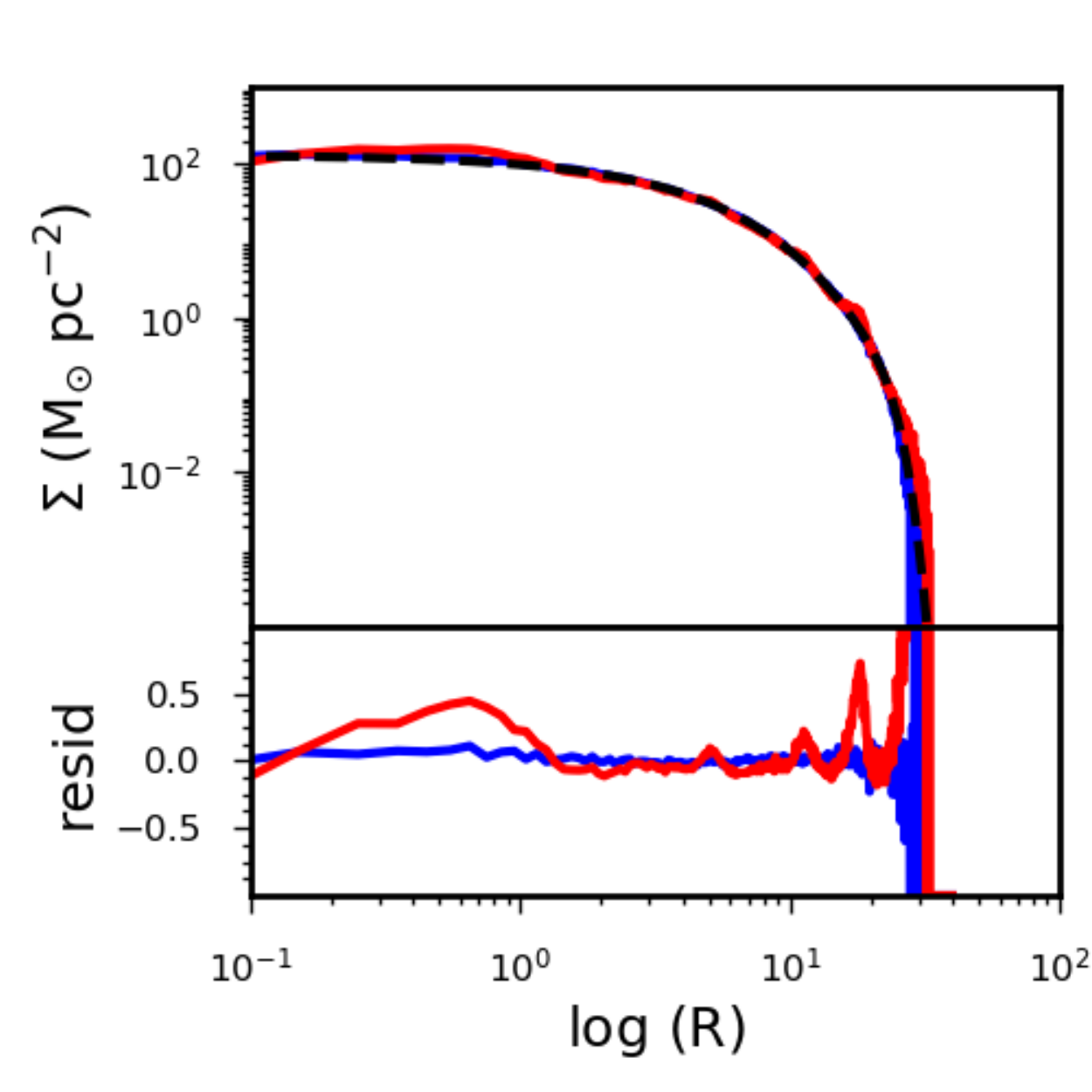} 
\caption{The azimuthally averaged surface density of the GalactICS-Gas4 gas \textcolor{black}{disc} as a function of radius (top) and the density residuals (bottom).  The dashed
black curve is the initial analytic profile and the solid blue and red lines are the surface densities
at $T=0$ and $1.0$ Gyr respectively.  The distance units are kpc.}
  \label{Fig:Gas4_SD}
\end{figure}
}

\section{A Milky Way Model}\label{Sec:MWModel}

In this section, we demonstrate the utility of
our new \textsc{GalactICS} code by generating and evolving
a realistic five-component model for a MW-like galaxy. Our model
is based on the mass models from \citet{McMillan2017}.
In that paper
observations of Maser line-of-sight velocities, 
the proper motion of Sgr A*, the terminal
velocity curve, the local vertical force, and
the total mass within 50 kpc were used to constrain the 
parameters of an analytic model for the mass and gravitational
potential of the Milky Way. The \citet{McMillan2017}
model comprised an NFW halo, 
thin and thick stellar \textcolor{black}{discs}, a bulge based
on \citet{Bissantz2002},
and both a HI and \textcolor{black}{H$_{2}$} gas \textcolor{black}{discs}. 

In this paper, we build a 
\textsc{GalactICS} model that uses the
"most likely" parameters from \citet{McMillan2017}.
We assume an NFW halo with
$\sigma_{h}=298~\kms$ 
and $r_{h}=19.6~\textrm{kpc}$ ($\rho_0 = 8.55\times 10^6\,M_\odot\,{\rm kpc}^{-3}$).
For the thin \textcolor{black}{disc} we assume
$M_{d,{\rm thin}}=3.52\times 10^{10} ~M_{\odot}$,
$R_{d,{\rm thin}}=2.5~\textrm{kpc}$ and $z_{d,{\rm thin}}=0.28~\textrm{kpc}$
while for the thick \textcolor{black}{disc} we assume
$M_{d,{\rm thick}}=7.2\times 10^{9} ~M_{\odot}$,
$R_{d,{\rm thick}}=3.02~\textrm{kpc}$, $z_{d,{\rm thick}}=0.83~\textrm{kpc}$.
Note that the scale height parameters are somewhat smaller than those
used in \citet{McMillan2017} to account the fact that the vertical structure
in those models is exponential in $|z|$ whereas
\textsc{GalactICS} \textcolor{black}{discs} have an approximately ${\rm sech(z)}^2$ form. The latter form
naturally arises from the assumption that the \textcolor{black}{discs} are vertically isothermal.

By design \textsc{GalactICS} bulges have a surface density profile that
is given, to a good approximation by the S\'{e}rsic profile.
To convert the \citet{Bissantz2002} bulge from \citet{McMillan2017} to a S\'{e}rsic one
we performed a simple parameter search using the 
inner 2 kpc of the density profile.
This search yielded $n\simeq 2$ for the S\'{e}rsic index, $R_b \simeq 0.64$ kpc
for the radial scale length, and 
and $\sigma_{b}=304~\kms$ for the velocity scale (see \citet{Widrow2008}).

\begin{figure}
\centering
    \includegraphics[width=80mm]{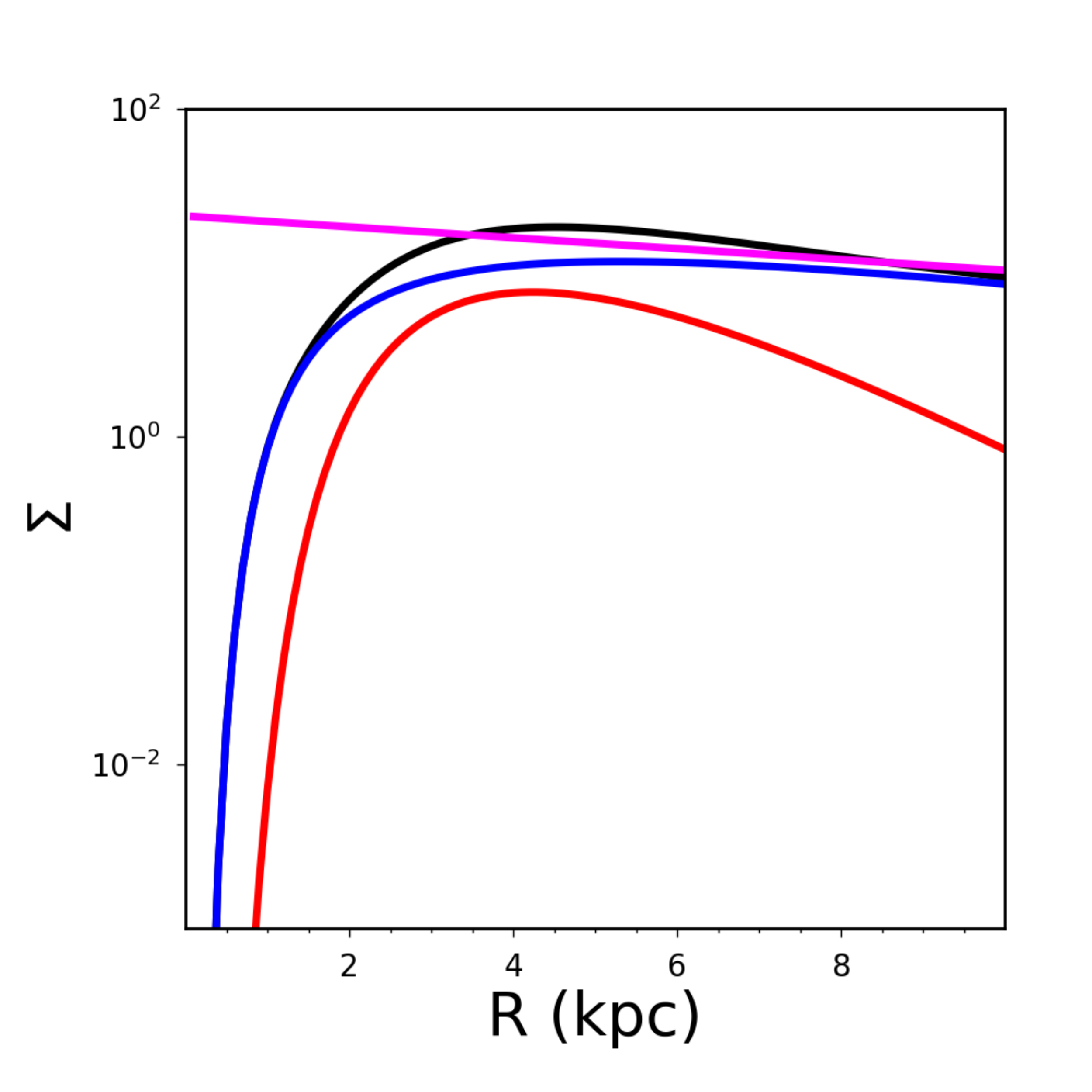} 
\caption{The two holed gas \textcolor{black}{discs} found by \citet{McMillan2017}  compared 
to our best fitting exponential gas \textcolor{black}{disc} (magenta).  The blue and red curves are the 
HI and H$_2$ gas \textcolor{black}{discs} and the black curve is the total gas surface density.
The surface density is in units of $M_{\odot}~\textrm{pc}^{-2}$.}
  \label{Fig:discComp}
\end{figure}

Finally, we consider the gas components from \citet{McMillan2017}. Since 
our current version of \textsc{GalactICS} has just a single exponential gas \textcolor{black}{disc}
we fit this model to the total gas surface density from the combined HI and \textcolor{black}{H$_{2}$}
gas \textcolor{black}{discs} of \citet{McMillan2017}.  Fig.
\ref{Fig:discComp} shows the comparison of the \textsc{GalactICS}
\textcolor{black}{disc} to the two \textcolor{black}{discs} in \citet{McMillan2017}.
Our \textcolor{black}{disc} has $M_{g}=2.39\times10^{10}~M_{\odot}$ 
and $R_{g}=13.1~\textrm{kpc}$. Note that the \citet{McMillan2017} 
\textcolor{black}{discs} have a hole in the centre so the structure of the gas components
in the two models is rather different though the total mass in gas
is very similar. 
Our model is constructed with a gas temperature of
$10^{4}~K$.

As with the models in the previous section, the system
is evolved using \textsc{Gadget-2}. Our N-body model
comprises $10^{6}$ gas particles, 
$5\times 10^{5}$ bulge particles, 
$2\times 10^{6}$ thin \textcolor{black}{disc} particles, 
$10^{6}$ thick \textcolor{black}{disc} particles, and $5\times 10^{6}$
halo particles. The system is evolved for a total of 2 Gyr.

\begin{figure}
\centering
    \includegraphics[width=80mm]{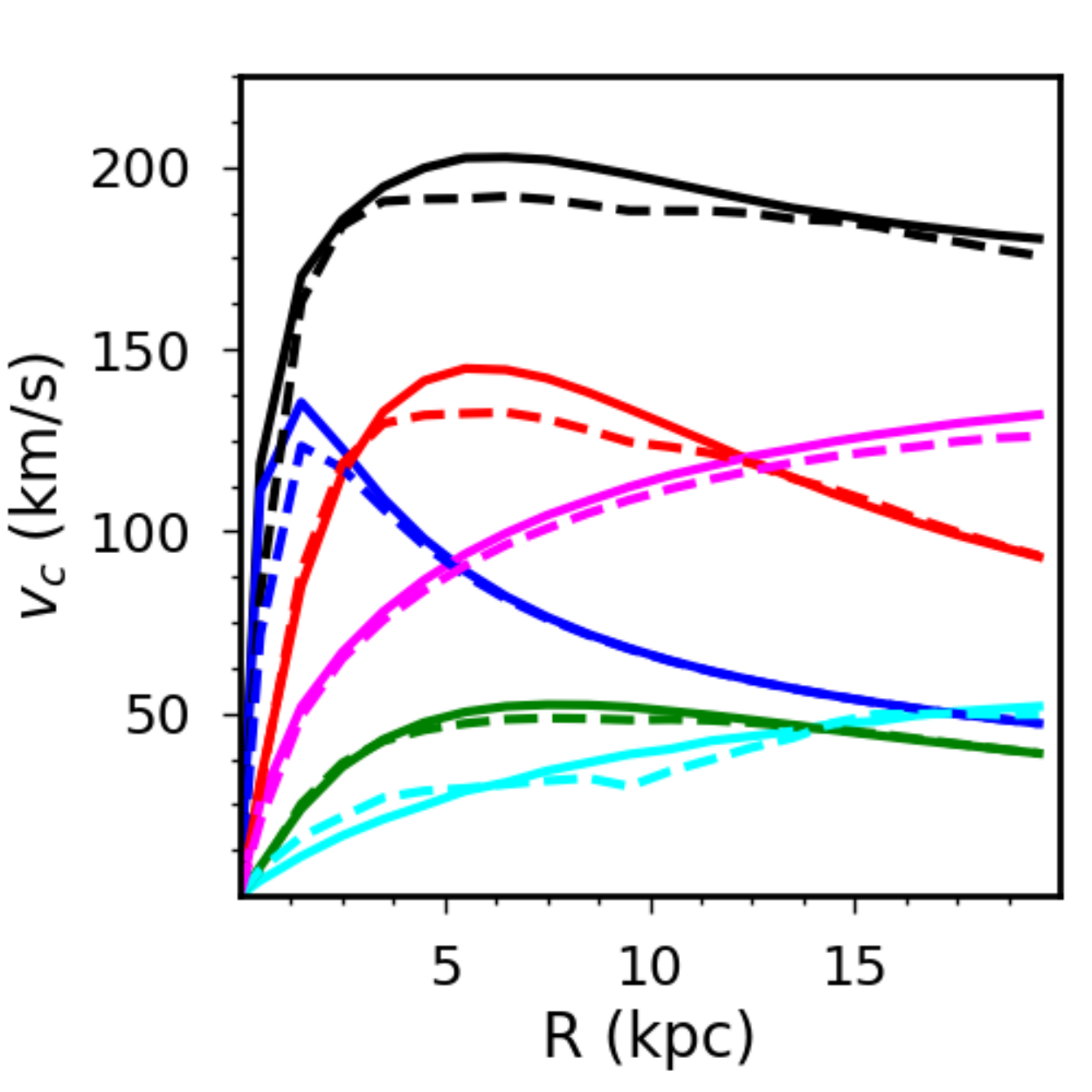} 
\caption{The rotation curve of our MW model at 
$T=0$ (solid lines) and $T=1$ Gyr (dashed lines).  The
black, blue, red, green, cyan, and magenta lines are the total 
curve and the contribution of bulge, thin \textcolor{black}{disc}, 
thick \textcolor{black}{disc}, gas \textcolor{black}{disc}, and halo respectively.}
  \label{Fig:MWRC}
\end{figure}

\begin{figure}
\centering
    \includegraphics[width=80mm]{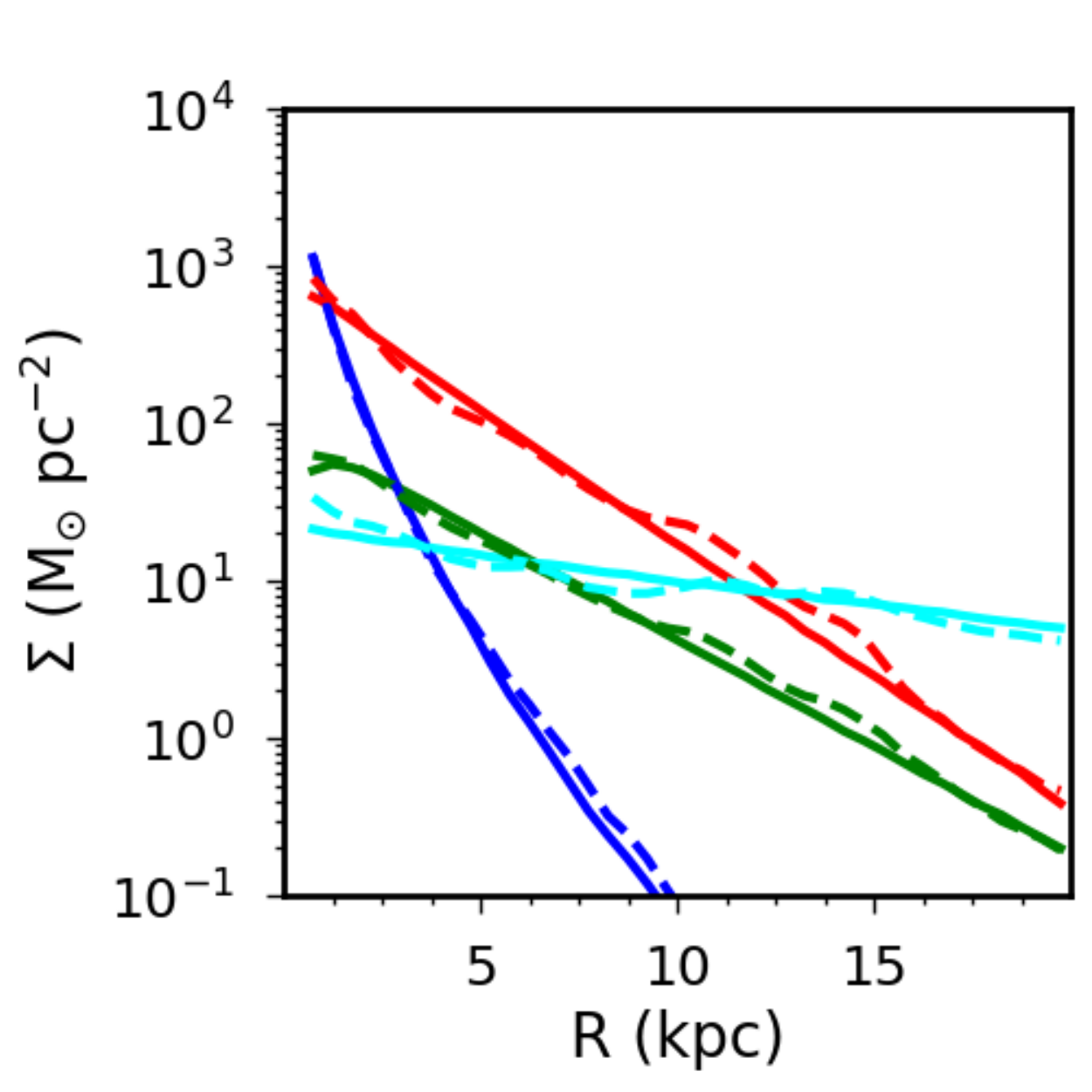} 
\caption{The surface density profiles of the four 
MW bayonic components at $T=0$ (solid lines) and $T=1$ Gyr (dashed
lines).  The bulge,
thin \textcolor{black}{disc}, thick \textcolor{black}{disc}, and gas \textcolor{black}{disc} are the blue, red, green, and 
cyan lines respectively.}
  \label{Fig:MWSD}
\end{figure}

Fig. \ref{Fig:MWRC} shows circular speed decomposition of our model
at $T=0$ and $T=1$ Gyr.
We see that the \textcolor{black}{disc} components (primarily the thin \textcolor{black}{disc}) dominate the radial force at radii of
about two \textcolor{black}{disc} scale lengths. Thus, we expect the model to form a bar
and spiral structure, which indeed it does. The surface density profiles
of the different components are shown in Fig. \ref{Fig:MWSD}. The thin \textcolor{black}{disc}
dominates the baryon mass budget at radii $R<13\,{\rm kpc}$ except in the 
innermost region where the bulge makes a comparable contribution to the 
surface density. On the other hand, the gas \textcolor{black}{disc}, which in this model
has a larger radial scale length, dominates the outer \textcolor{black}{disc}.

The evolution of the circular speed and azimuthally averaged 
surface density profiles is due to the formation of a bar
and spiral arms.  From the surface density profiles
we see that mass in the \textcolor{black}{disc} components
moves outward from $R\sim 3-5$ kpc to $R\sim 9-11$ kpc.
This mass redistribution leads to a change in the shape
of the rotation curve.

\begin{figure*}
\centering
    \includegraphics[width=\textwidth]{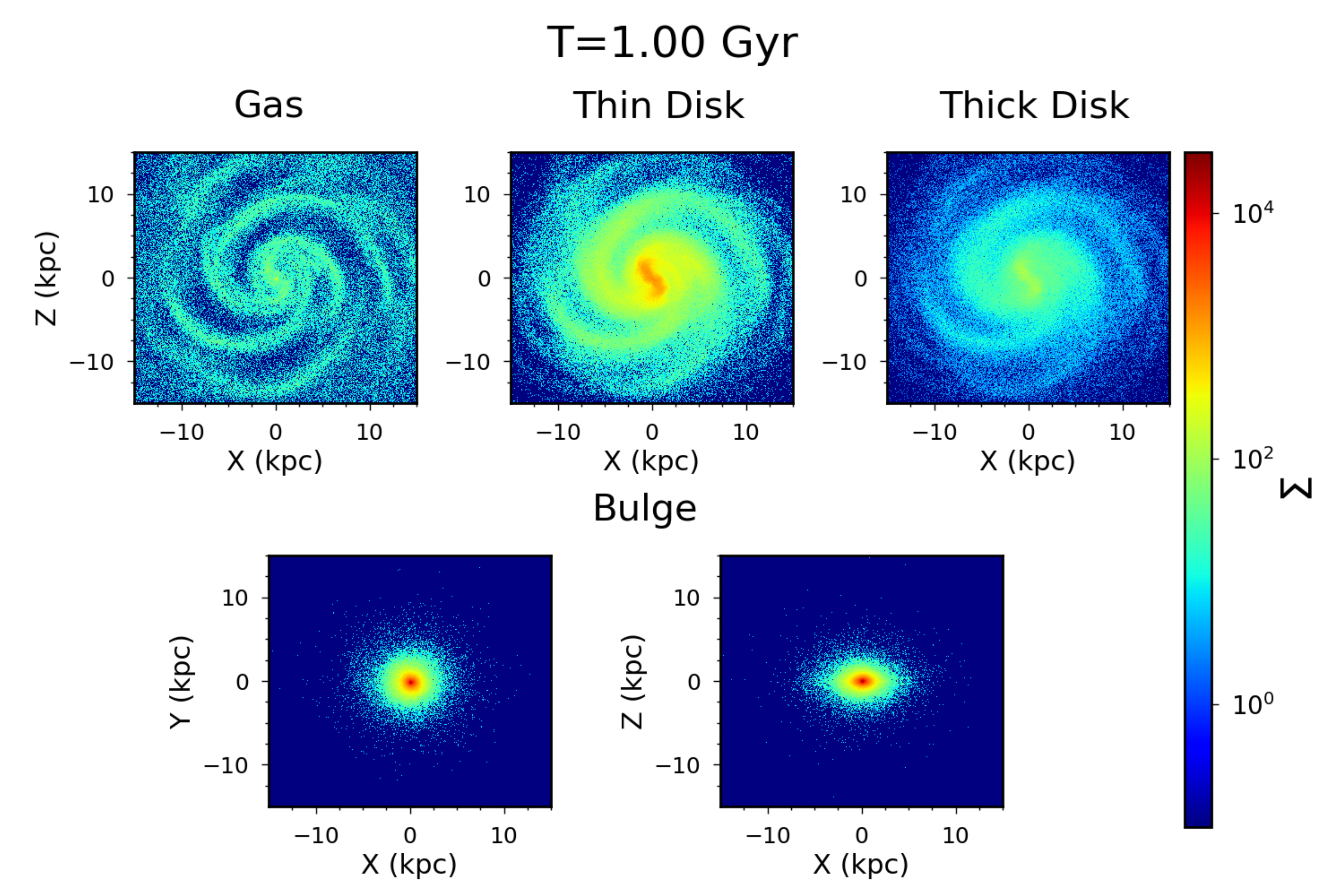} 
\caption{The $X-Y$ surface density for the gas \textcolor{black}{disc}, thin and thick \textcolor{black}{discs},
and the $X-Y$ and $X-Z$ surface density of the bulge.
The surface density is in units of $M_{\odot}~\textrm{pc}^{-2}$.}
  \label{Fig:MWSDMap}
\end{figure*}

\begin{figure*}
\centering
    \includegraphics[width=\textwidth]{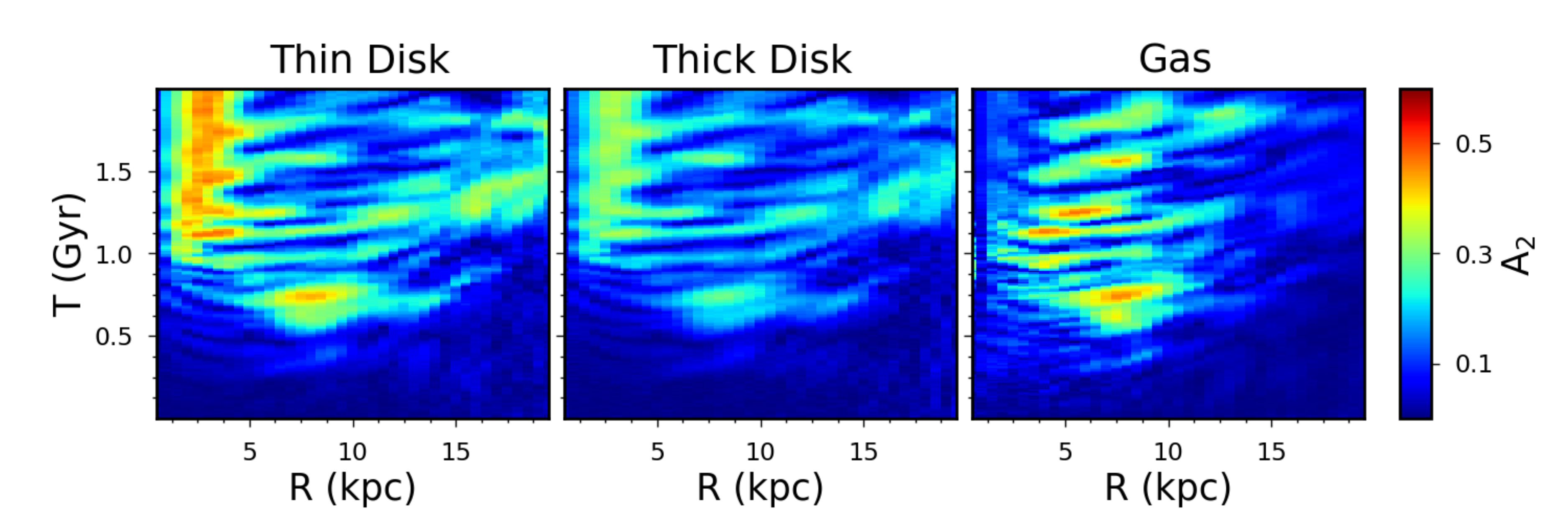} 
\caption{The $A_{2}$ moment as a function of radius and time for the 
thin, thick, and gas \textcolor{black}{discs}.}
  \label{Fig:BarRadEvolve}
\end{figure*}

Figure \ref{Fig:MWSDMap} shows a snapshot
of the model at $T=1$ Gyr. As noted above, the \textcolor{black}{disc} components have all developed
bar and spiral structure, although the bar strength 
varies considerably between the various \textcolor{black}{disc} 
components, \textcolor{black}. The difference between the components
can be seen more clearly in Fig. \ref{Fig:BarRadEvolve},
which shows the evolution of $A_{2}$ for
the three \textcolor{black}{disc} components as a function of radius and
time where $A_2$ is the amplitude of the $m=2$ Fourier
component for the surface density where $m$ is the usual azimuthal
mode number.
At all times and radii, the 
thin and thick \textcolor{black}{discs} show similar structure, although 
the $A_{2}$ moment is always weaker in the 
thick \textcolor{black}{disc}.  Both stellar \textcolor{black}{discs} show a clear 
bar component with peaks in the $A_{2}$ moment at 
$R\approx 2.5 $ kpc for \textcolor{black}{$T\ge 1.0 $} Gyr.  The gas \textcolor{black}{disc} 
has a much more transitory structure. While there is 
some structure in the bar region, the dominant $m=2$
structures are at larger radii and clearly correspond
to the two-armed spiral structure seen in Fig. \ref{Fig:MWSDMap}.

\textcolor{black}{
The difference in the bar strengths between the thin and
thick discs is similar what is observed in both 
nature and simulations.
\citet{Athanassoula1983} ran simulations of 
galaxies with single discs and noted that 
dynamically colder discs formed thinner
bars than hotter discs.  
\citet{Athanassoula2003} followed this 
investigation with a theoretical and numerical
exploration of the relationship between angular momentum
transfer and bar properties.  In that work it was 
found that increased disc velocity dispersions inside
the co-rotation radius decreased the bar strength.
More recently \citet{Athanassoula2016} ran simulations 
of merging galaxies.  They found that the thin disc also
formed a stronger bar component than the thick disc.  The bar
present in their thick discs have the same length and
orientation as in the thin disc bar, but have larger 
vertical extents and are significantly more oval in 
nature.  \citet{Athanassoula2018} found similar results
using a more advanced analysis of these 
merging simulations clearly show the difference
between the thin and thick discs.
The simulations of \citet{Bekki2011}
examined the formation of the bulge from a 
single or two-disc scenario.  They found that
a thick disc is less influenced by the presence of 
a bar, allowing for a steep vertical metallicity gradient.
Similarly, \citet{Debattista2017} utilized variety
of different simulations to show that populations with 
lower radial velocity dispersions form stronger bars
than those that are dynamically hotter. \citet{Fragkoudi2017}
found similar results in their
simulations, noting that the thin disc
bar was $\sim$50$\%$ stronger than the bars found in the thick
disc, as well as being significantly more elongated.
}

The bar and spiral structure seen in our evolved
model are similar to what are observed in the Milky Way.
For example, the bar in the MW has a length of about $3-4$ kpc
(see \citet{Bissantz2002} and references therein), which is
comparable to what we find in our simulation. Likewise,
the spiral structure shows a strong two-armed pattern with
evidence for weaker four or more armed spiral structure.
Our structures do seem to have a tighter winding pattern
than what is seen in the Galaxy. 

The origin of the bar and spiral structure pattern of the 
Milky Way is still an open question. For example, \citet{Purcell2011}
argued that these structures might be the result of an encounter
between the \textcolor{black}{disc} and the Sagittarius dwarf. Our results suggest
that the \textcolor{black}{disc} can form structures of this type through internal
instabilities. \textcolor{black}{ However, this simulation has only been
run for a few orbits.  Further investigation must be done to examine
whether the bar and spiral structure seen here will persist
over dozens of dynamical times.}

\section{Summary and Conclusions}\label{Sec:Con}

In this paper, we have introduced a new version of the \textsc{GalactICS} 
code that is able to 
generate equilibrium models for a galaxy consisting of thin and thick
stellar \textcolor{black}{discs}, a stellar bulge, 
a dark halo, and gas \textcolor{black}{disc}. The models can be used to 
generate ICs for N-body simulations that, in turn, can be used to study the
dynamics of real galaxies and, in particular, the formation
of bars and spiral structure.

\textcolor{black}{Test bed simulations of models with just a gas and dark
halo showed the ability of GalactICS to produce both
stable and unstable models.  A transient is 
present in some gas-halo only models.  The 
inclusion of a stellar \textcolor{black}{disc} suppresses these
transients, yielding stable models.  Regardless 
of the presence of a transient, the 
azimuthally averaged surface densities remain
roughly constant.  The tangential velocities
of the gas particles have an increase in 
random motions, but the gravitationally 
calculated circular speed remains 
constant.}

We demonstrated the applicability of the model by constructing
a multi-component {\it dynamical model} for a MW-like galaxy based
on the {\it mass model} of
\citet{McMillan2017}.  We found that this model produces a 
relatively strong bar and spiral structure by 
\textcolor{black}{$T=1.0$ Gyr}. The bar is present in all of the 
\textcolor{black}{disc}-like baryonic components though strongest in the thin \textcolor{black}{disc}.
Spiral arms are also present in all of the \textcolor{black}{disc} components
though noticeably thinner and more prominent in the gas \textcolor{black}{disc}.

\textsc{GalactICS} provides an excellent tool for dynamical studies of 
galaxies. In particular, it generates N-body ICs that 
that can be used to study the dynamics of observationally-motivated
galaxy models. The inclusion of a gas 
\textcolor{black}{disc} opens up new and exciting possibilities that we
are eager to explore in future work.
Indeed, the models have already been used to study the rotation
curves of barred spiral galaxies \citep{Toky2018}.
Many standard algorithms for galaxy modelling fail for 
particular bar orientations. A suite of tailored ICs generated
with \textsc{GalactICS} allowed us to model galaxies of this type.
The simulations shown in the present work suggest that 
dynamical simulations combined with observations might allow one
to constrain the models in a fashion not possible with 
mass models of the type considered in \citet{McMillan2017}.

 CC's work is based upon research supported by the South African Research Chairs Initiative (SARChI) 
 of the Department of Science and Technology (DST), the SKA SA and the
 National Research Foundation (NRF). 
 ND's  work is supported by a SARChI's South African SKA Fellowship. 
 LMW is supported by the Natural Sciences and Engineering 
 Research Council of Canada through Discovery
Grants.  The numerical simulations were performed at the
Centre for High Performance Computing.

\bibliographystyle{astron}

\bsp	
\label{lastpage}
\end{document}